\documentclass[epsf,prb,twocolumn,showpacs,nofootinbib]{revtex4-1}

\usepackage[pdftex]{graphicx}
\usepackage{dcolumn}
\usepackage{bm}
\usepackage{epsfig}
\usepackage{latexsym}
\usepackage{amsmath}
\usepackage{amsfonts}
\usepackage{amssymb}
\usepackage{float}
\usepackage{caption}
\usepackage{graphicx}
\graphicspath{{images/}}
\usepackage{subcaption}
\usepackage{color}
\usepackage{array}
\usepackage{framed}

\begin{document}

\title{Fractonic line excitations : an inroad from 3d elasticity theory}
\author{Shriya Pai and Michael Pretko}
\affiliation{Department of Physics and Center for Theory of Quantum Matter, University of Colorado, Boulder, CO 80309, USA}
\date{\today}

\begin{abstract} 
We demonstrate the existence of a fundamentally new type of excitation, \textit{fractonic lines}, which are line-like excitations with the restricted mobility properties of fractons. These excitations, described using an amalgamation of higher-form gauge theories with symmetric tensor gauge theories, see direct physical realization as the topological lattice defects of ordinary three-dimensional quantum crystals. Starting with the more familiar elasticity theory, we show how it maps onto a rank-4 tensor gauge theory, with phonons corresponding to gapless gauge modes and disclination defects corresponding to line-like charges.  We derive flux conservation laws which lock these line-like excitations in place, analogous to the higher moment charge conservation laws of fracton theories. This way of encoding mobility restrictions of lattice defects could shed light on melting transitions in three dimensions.  This new type of extended object may also be a useful tool in the search for improved quantum error-correcting codes in three dimensions.
\end{abstract}
\maketitle

\normalsize

\tableofcontents

\section{Introduction}

\textit{Overview}: Quantum phases of matter have attracted interest from several disciplines such as condensed matter theory, quantum information and quantum materials. For describing phases like quantum Hall states\cite{fqh1,fqh2} or quantum spin liquids\cite{sl1,sl2}, the framework of gauge theory offers a natural description.  More recently, quantum phases of matter that host immobile quasiparticle excitations were discovered, and this led to the birth of the fracton phenomenon \cite{fractonreview}. Although fractons are completely immobile in isolation, there are models that host excitations which have restricted mobility only in certain directions: these are subdimensional particles. While fractons were originally conceived in exactly solvable lattice models in three dimensions \cite{chamon,bravyi,cast,haah,yoshida,haah2,fracton1,fracton2}, these particles were found to have a description in terms of higher-rank tensor gauge theories \cite{sub,genem,alex,cenke}, with conservation laws involving higher moments of charges, accounting for their immobility. Much progress has ensued from this understanding, with developments addressing finite-temperature effects \cite{screening,glassy}, gravitational behavior \cite{mach}, connections with quantum Hall physics\cite{theta,matter} and deconfined quantum criticality\cite{deconfined}, and many other developments. \cite{williamson,hanlayer,sagarlayer,parton,slagle,bowen,nonabel,balents,field,valbert,correlation,simple,entanglement,bernevig,albert,generic1,generic2,z3,sspt,higgs1,higgs2,foliated} 

Very recently, a direct realization of fractons in two-dimensional crystals was identified \cite{elasticity}, where two-dimensional elasticity theory was mapped onto a tensor gauge theory with fractonic excitations.\cite{footnote}  This duality is the tensor analogue of the more conventional particle-vortex duality. The excitations of two-dimensional elasticity theory were shown to be in one-to-one correspondence with those of a rank-2 $U(1)$ tensor gauge theory, specifically the scalar charge theory studied in Reference \onlinecite{sub}.  The disclination and dislocation defects map onto fractonic charges and dipoles, respectively, while phonons map onto the gapless gauge modes.  Additionally, some predictions about the phases and phase transitions of the fracton gauge theory were  tallied with established results in elasticity theory. On the other hand, the conservation laws of fracton gauge theories provide a systematic tool for analyzing the dynamics of lattice defects.  Subsequent work has also proposed that a chiral rank-2 gauge theory may describe an exotic topological elasticity theory.\cite{gromov}

\textit{Summary of results}: In the present work, we extend this mapping between the elastic theory of quantum crystals and a fractonic tensor gauge theory of the appropriate rank to three dimensions, and thereby uncover a new kind of excitation: \textit{fractonic lines}. The physics of this theory borrows from both antisymmetric tensor gauge theories (that host line excitations) and symmetric tensor gauge theories (that host fractons). In neat correspondence with the 2d version of the duality, we demonstrate how the mobility restrictions of lattice defects are encoded in conservation laws of the dual gauge theory.  We formulate conservation laws on the \emph{flux} of lines through two-dimensional surfaces, in a natural generalization of the usual higher moment charge conservation laws of fracton theories. The conserved charges and dipoles of 2d scalar charge theory transform into \textit{conserved fluxes} and \textit{dipoles of fluxes} in this rank-4 tensor gauge theory. These conservation laws serve to immobilize the line excitations, locking them in place along fixed one-dimensional paths.  We note that this is phenomenologically similar to localization in the context of line-like objects\cite{extended}, but in the present case, immobility arises in the absence of any disorder.

In this way, the notion of fracton behavior is extended from point particles to line-like objects, representing a fundamentally new class of excitation.  These fractonic lines have already seen direct physical realization in ordinary three-dimensional crystals.  In this context, the language of fractons provides a convenient way of encoding the mobility restrictions of topological lattice defects, potentially providing insight into three-dimensional melting transitions.  Our work also opens the door for investigations into other types of line-like excitations with fractonic or subdimensional behavior.  In particular, it seems plausible that fractonic line excitations could be realized in stabilizer code models describing gapped phases of matter.  Since quantum codes with line-like excitations have better error-correcting properties than those with all point-like excitations, fractonic lines may also provide a useful ingredient in the search for more robust quantum error-correcting codes.

\section{Duality}

A crystal is characterized by a lattice phonon displacement field $u^{i}$, which is the Goldstone mode of the crystal corresponding to spontaneous breaking of translational symmetry. Using the fact that the strain is the difference between the distortion-induced metric and the undistorted metric, the linear part of the symmetrized strain tensor may be written as\cite{chaikin,landau} 
\begin{equation}
u_{ij} = \dfrac{1}{2}(\partial_{i}u_{j}+\partial_{j}u_{i})
\end{equation} 
The most general linearized low-energy action for the lattice displacement is given by
\begin{equation}
S = \int d^{3}x dt \dfrac{1}{2}[(\partial_{t}u^{i})^{2}-C^{ijkl}u_{ij}u_{kl}],
\end{equation}
where $C^{ijkl}$ is a tensor of elastic coefficients.  We now separate the displacement field into singular and smooth phonon pieces, allowing us to write the strain tensor as $u_{ij} = u^{(s)}_{ij} + \dfrac{1}{2}(\partial_{i} \widetilde{u}_{j} + \partial_{j} \widetilde{u}_{i})$, where $\widetilde{u}_{i}$ is a smooth function. The singular part of the strain tensor $u^{(s)}_{ij}$ arises from having a singularity in the bond angle. We note that the three-dimensional description requires a \textit{vector bond angle} whose derivative is given by 
\begin{equation}
\partial_{i}\theta_{\textrm{bond}}^{j} = \epsilon^{jkl}\partial_{k}u_{il}^{(s)}
\end{equation}
Note that, strictly speaking, the angular variable $\theta^j$ itself is not a true vector, due to the non-commutativity of rotations.  Nevertheless, \emph{derivatives} of $\theta^j$, defined in terms of infinitesimal differences of angles, transform as valid tensor objects.  The disclination density is then $s^{ij} = \epsilon^{ikl}\partial_{k}\partial_{l}\theta_{\textrm{bond}}^{j}$, which arises from the singular part of $u_{ij}$ via 
\begin{equation}
s^{ij} = \epsilon^{ikl}\epsilon^{jmn}\partial_{k}\partial_{m}u_{ln}^{(s)}
\label{disclination}
\end{equation}
We rewrite the action in equation $(2)$ by introducing Hubbard-Stratonovich fields $\sigma_{ij}$ and $\pi_{i}$, which correspond to the symmetric stress tensor and the lattice momentum respectively. The action is then
\begin{multline}
S = \int d^{3}x dt \bigg[\dfrac{1}{2} C^{-1}_{ijkl}\sigma^{ij}\sigma^{kl} - \dfrac{1}{2}\pi^{i}\pi_{i} - \sigma^{ij}(\partial_{i}\widetilde{u}_{j}\\ + u_{ij}^{(s)})+\pi^{i}\partial_{t}(\widetilde{u}_{i}+u^{(s)}_{i})\bigg]
\end{multline}
The action is now linear in the smooth function $\widetilde{u}_{i}$, which can be integrated out to give the constraint
\begin{equation}
\partial_{t}\pi^{i}-\partial_{j}\sigma^{ij}=0,
\end{equation}
which is Newton's equation of motion for the lattice. In order to solve this constraint, we first introduce rotated fields,
\begin{equation} 
\begin{aligned}
\pi^{i} &= \epsilon^{iab}B_{ab}\\
\sigma^{ij} &= \epsilon^{ikl}\epsilon^{jmn}E_{klmn}^{\sigma},
\end{aligned}
\end{equation}
where $E_{klmn}^{\sigma}$ is symmetric under $(kl) \leftrightarrow (mn)$ and antisymmetric under $(k \leftrightarrow l)$ and $(m \leftrightarrow n)$.  With this definition, the above equation of motion rotates into generalized Faraday's equation \cite{genem}, 
\begin{equation}
\partial_{t}B_{ab} - \epsilon^{jmn}\partial_{j}E_{abmn}^{\sigma}=0
\end{equation}
In the static limit, we have a magnetostatic constraint on the electric field, $\epsilon^{jmn} \partial_{j} E_{abmn}^{\sigma}=0$, from which we can see that either the third or the fourth index of $E_{abmn}^{\sigma}$ should have a derivative in it.  This allows us to write $E_{abmn}^{\sigma}$ in terms of a symmetric rank-2 tensor electric potential $\phi_{ij}$ as follows:
\begin{equation}
E_{abmn}^{\sigma}=\partial_{m}\partial_{b}\phi_{an}-\partial_{m}\partial_{a}\phi_{nb}+\partial_{a}\partial_{n}\phi_{bm}-\partial_{b}\partial_{n}\phi_{ma}
\end{equation}
Relaxing the magnetostatic constraint, we can complete the analogy with Maxwell theory by writing the electric and magnetic tensors in terms of $\phi$ and a rank-4 tensor potential $A_{ijk\ell}$ as
\begin{multline}
\begin{aligned}
B_{ab} &= \epsilon^{jmn} \partial_{j} A_{abmn}\\
E_{abmn}^{\sigma} &= \partial_{m}\partial_{b}\phi_{an}-\partial_{m}\partial_{a}\phi_{nb}+\partial_{a}\partial_{n}\phi_{bm}\\ & -\partial_{b}\partial_{n}\phi_{ma} - \partial_t A_{abmn}
\end{aligned}
\end{multline}
where $A_{abmn}$ has the same index symmetries as $E_{abmn}$.  The fields $E_{abmn}^{\sigma}$ and $B_{ab}$
are invariant under the generalized gauge transformation on the potentials,
\begin{multline*}
\begin{aligned}
 A_{abmn} & \rightarrow A_{abmn} + (\partial_{b}\partial_{m}\alpha_{na}-\partial_{m}\partial_{a}\alpha_{nb}+\partial_{n}\partial_{a}\alpha_{bm}\\ &-\partial_{b}\partial_{n}\alpha_{ma})\\\ 
\phi_{ab} & \rightarrow \phi_{ab} + \partial_t \alpha_{ab} 
\end{aligned}
\end{multline*}
for some tensor $\alpha_{ab}(\vec{x},t)$ with arbitrary spacetime dependence. We now express the action in equation $(2)$ in the language of the rank-4 gauge theory;  using equations $(11)$ and $(4)$ and integrating by parts, we obtain,
\begin{multline}
S = \int d^{3}x dt \bigg[\dfrac{1}{2} \widetilde{C}^{-1}_{ijklpqrs}
E^{ijkl}_{\sigma}E^{pqrs}_{\sigma} - \dfrac{1}{2}B^{ij}B_{ij} - \rho^{ij}\phi_{ij}\\ - J^{ijkl}A_{ijkl}\bigg],
\end{multline}
where $J^{ijkl}$ is the rank-4 current tensor describing the motion of dislocations and disclinations \cite{genem,interstitials}, and $\widetilde{C}^{ijklpqrs} = \epsilon^{aij}\epsilon^{bkl}\epsilon^{cpq}\epsilon^{drs}
C_{abcd}$ is a redefined tensor of elastic coefficients. The action in equation $(12)$ resembles that encountered in \onlinecite{elasticity}, albeit with tensor charges playing the role of disclinations.

In order to obtain the Gauss's law for this theory, we write the Hamiltonian by introducing the canonical conjugate electric tensor $E_{ijkl} = \partial \mathcal{L}/ \partial \dot{A}^{ijkl} = \widetilde{C}^{ijklpqrs}E_{pqrs}^{\sigma} $ as follows: 
\begin{multline}
\mathcal{H} = \int d^{3}x \bigg( \dfrac{1}{2}\widetilde{C}^{ijklpqrs}E_{ijkl}E_{pqrs} + \dfrac{1}{2}B^{ij}B_{ij} +\rho^{ij}\phi_{ij}\\ + A^{ijkl}J_{ijkl}\bigg)
\end{multline}
Note the distinction between the Hamiltonian variable $E_{ijk\ell}$ and the Lagrangian variable $E^\sigma_{ijk\ell}$, which differ by a tensor factor.  The potential $\phi_{ij}$ acts as a Lagrange multiplier and imposes the Gauss's law constraint
\begin{equation}
 \partial_{i}\partial_{k}E^{ijkl} = \rho^{jl},
 \label{gauss}
\end{equation}
where $\rho^{jl}$ is symmetric. Furthermore, we see that the constraint
\begin{equation}
\partial_{j}\rho^{jl}=0
\end{equation} 
follows from equation \ref{gauss}.  A direct analogy with $\vec{\nabla}\cdot\vec{B} = 0$ from Maxwell theory or $\partial_{i}\rho^{i}=0$ from the 2-form gauge theory (see Appendix A) forces the charges of the rank-4 theory into \textit{extended objects}.

From the Hamiltonian, we can also derive the generalized Ampere's equation to be
\begin{equation}
\dfrac{1}{2}(\epsilon^{akl}\partial_{a}B^{ij} + \epsilon^{aij}\partial_{a}B^{kl}) + J^{ijkl} + \partial_{t}E^{ijkl} = 0.
\end{equation}
By applying $\partial_i\partial_k$ to this equation, we obtain
\begin{equation}
\partial_{t}\rho^{jl} + \partial_{i}\partial_{k} J^{ijkl} = 0,
\end{equation}
which is the continuity equation for the theory.

\section{Mobility restrictions and conservation laws}

In analogy with two-dimensional fracton-elasticity duality, we expect that the topological defects of three-dimensional elasticity theory, and hence the gauge charges of the rank-4 tensor gauge theory, will have mobility restrictions similar to those of fractons.  However, it is not immediately obvious what it means for a line-like excitation to behave as a fracton.  In order to formulate the concept of a fractonic line, we must first revisit the key physics leading to the immobility of fractons.

\begin{figure}[t!]
 \centering
 \includegraphics[scale=0.37]{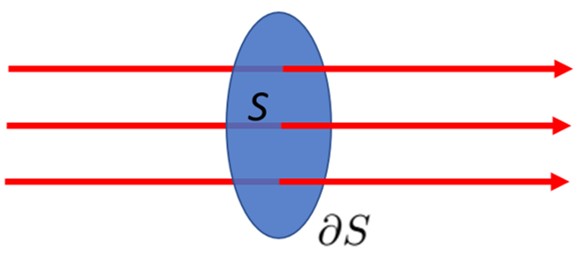}
 \caption{To formulate conservation laws for line-like excitations, we examine the \emph{flux} of lines through an arbitrary two-dimensional surface $S$.}
 \label{fig:flux}
 \end{figure}

Fractons are immobile due to the presence of higher moment charge conservation laws, such as dipole moment conservation.  We must therefore investigate what type of conservation laws apply to a line-like excitation.  The most direct approach is to examine the point-like \emph{flux} of these lines through two-dimensional surfaces, as depicted in Figure \ref{fig:flux}.  We can then formulate more familiar conservation laws, such as charge conservation, in terms of these flux points.  Such conservation laws are already present even in more familiar two-form gauge theories, as reviewed in Appendix A.  For our rank-4 theory, we will see that the line-like excitations exhibit conservation of \emph{dipole moment} of flux on an arbitrary cross-section.  This conservation law will restrict the flux points from moving around on the two-dimensional surface.  Since we can choose any cross-section, every point of the line becomes locked in place, freezing the entire line along a specific one-dimensional path.  In this way, the concept of a fracton is generalized to extended objects, giving us fractonic line excitations.

\subsection{Gauge charges as fractonic lines}

We begin by examining the flux of our tensor charge, defined in Equation \ref{gauss}, through an arbitrary two-dimensional surface $S$.  The flux due to the tensor charge is defined as
\begin{equation}
f^{i} = \int_{S} \rho^{ij} dn_{j}
\end{equation}
Our proof of conservation of this flux will closely follow the analysis of a two-form gauge theory, discussed in Appendix A.  We begin by introducing a rotated electric field $V_{mn}$
\begin{equation}
E^{ikjl} = \epsilon^{ikm}\epsilon^{jln}V_{mn},
\label{rotated}
\end{equation}
In terms of this variable, the flux through a surface $S$ becomes 
\begin{equation}
f^{i} = \int_{S} dn_{j}\epsilon^{ikl}\epsilon^{jlm} \partial_{k}\partial_{l}V_{mn}
\end{equation}
Using Stokes' theorem, we then find:
\begin{multline}
f^{i} = \int_{S} dn_{j} \epsilon^{jlm} \partial_{l} (\epsilon^{ikn}\partial_{k} V_{nm}) = \oint_{\partial S} (\epsilon^{ikn}\partial_{k} V_{nm}) ds^{m}\\ = \textrm{boundary term} 
\end{multline}
We see that the flux through the two-dimensional surface $S$ is encoded on the one-dimensional boundary $\partial S$.  Following the logic of Reference \onlinecite{sub}, we can then conclude that the flux $f^i$ is a locally conserved quantity on the surface $S$.  There is no local operator in the bulk of $S$ which can change the total flux, since any flux-changing operator must act on the boundary $\partial S$.

This bulk-boundary relationship for $f^i$ also allows us to identify the physical meaning of flux on the elasticity side of the duality.  At the boundary, away from charges, we can write $E^{ijkl} = \epsilon^{ijp}\epsilon^{klq}u_{pq}$, such that $V_{mn} = u_{mn}$.  Recalling that the change in bond angle takes the form $\partial_{i}\theta_{\textrm{bond}}^{j} = \epsilon^{jkl}\partial_{k}u_{il}$, we can write
\begin{align}
f^{i} = \oint_{\partial S} ds^m(\epsilon^{ikn} \partial_{k} u_{nm}) = \oint_{\partial S} ds^{m} \partial_{m} \theta^{i}_{\textrm{bond}} = \Delta \theta^{i}_{\textrm{bond}},
\end{align}
which indicates that the flux due to a tensor gauge charge excitation corresponds to a winding of the bond angle around the normal to the surface $S$.  In other words, the line-like excitations described by $\rho^{ij}$ correspond to the disclination defects of a 3d crystal.

In addition to the conserved flux associated with disclinations, this gauge theory also has an extra conservation law.  We define the \emph{dipole} of flux to be:
\begin{equation}
\mathcal{P}_{i}^{\textrm{flux}} = \int_{S}dn_{l}(x^{k} \rho^{jl} \epsilon_{kji}).
\end{equation}
Using Equation \ref{rotated}, we can write 
\begin{align}
\begin{split}
\mathcal{P}_{i}^{\textrm{flux}} &= \int_{S} dn_{l} \epsilon_{jik}\epsilon^{jma}\epsilon^{lnb}x^{k}\partial_{m}\partial_{n}V_{ab}\\
&= \oint_{\partial S} ds^{m} \big(\partial_{i}x^{k}V_{km} -\partial_{k}x^{k}V_{im} +V_{im}\big)\\
&= \textrm{boundary term},
\label{dipolebound}
\end{split}
\end{align}
By the same logic as the charge conservation law, this equation indicates that dipole of flux is a locally conserved quantity.  No local operator affecting the bulk of surface $S$ can create a flux dipole.  Rather, the total flux dipole can only change via operations at the boundary $\partial S$.

Starting with the boundary term in equation \ref{dipolebound},
we use the low-energy form of $E^{ijkl}$ and rearrange a few terms to obtain:
\begin{align}
\mathcal{P}_{i}^{\textrm{flux}} &=  \oint_{\partial S} ds_{m} \big[ \partial_{i} (x^{k}u_{km}) - \partial_{k} (x^{k}u_{im}) + u_{im} \big].
\end{align}
Assuming that there are no disclinations contained in $S$, so that the first two terms above do not wind non-trivially around $\partial {S}$, we can simplify this expression to
\begin{equation}
\mathcal{P}_{i}^{\textrm{flux}} =  \oint_{\partial S} ds^{m} \partial_{m} \big( \epsilon^{ik}u_{k} \big) = \epsilon^{ik}\Delta u_{i} = \epsilon^{ik}b_{k},
\end{equation}
where $b_{k}$ is a Burgers vector.  This indicates that line-like excitations carrying flux dipole correspond to dislocation defects, in agreement with the fact that dislocations are bound states of two disclinations.\cite{seung, chaikin, landau, sarang}  The Burgers vector (flux dipole moment) can either be in the direction of the line excitation, yielding a screw dislocation, or perpendicular to the direction of the line, yielding an edge dislocation.

From the flux dipole conservation law, we see that edge dislocations in the $z$ direction are made up of bound states of $\rho_{zz}$, while for a cubic lattice, screw dislocations in the $z$ direction are made up of bound states of $\rho_{xz}$ and $\rho_{yz}$, corresponding to the conserved quantum number $\int (y\rho_{xz} - x \rho_{yz})$.  To visualize screw dislocations, we can think of $\rho_{xz}$ and $\rho_{yz}$ as occurring on lattice plaquettes in the $XZ$ and $YZ$ planes respectively. But $\rho^{ij}$ in our theory is subject to an additional constraint, $\partial_{i} \rho^{ij} = 0$. To have a configuration of $\rho_{xz}$ and $\rho_{yz}$ consistent with this constraint, a screw dislocation can be formed as a small closed tube made from these plaquette variables.

\subsection{Glide constraint on dislocations}

In addition to the fractonic behavior of disclination lines, the edge dislocations of elasticity theory have a mobility constraint which must be accounted for in the gauge theory language.  Specifically, an edge dislocation can only move freely in the plane spanned by its Burgers vector and the direction of the dislocation line.  Meanwhile, perpendicular motion requires the creation of vacancy or interstitial defects.  To account for this restriction, we consider a component of the \textit{quadrupole flux}, which we define as follows:
\begin{equation}
\mathcal{Q}^{j}_{\textrm{flux}} =\int_{S} dn_{i} \rho^{ij} x^{2}.
\end{equation}
We now work in the special case where $S$ is a plane, such that the normal direction $dn_i$ is fixed ($e.g.$ $i=z$), and we also choose $j=i$, giving
\begin{multline}
\int_{S} dn_{i} (\rho^{ii}x^{2} - 2 E^{\,\,iki}_{k}) = \oint_{\partial S} ds^{n} (x^{2} \epsilon_{ikn} \partial_{l} E^{ikil}\\ - 2x_{k} \epsilon_{iln}E^{ikil}), 
\label{quadcons}
\end{multline}
(where there is no sum over $i$).  Since this particular component of the quadrupole flux can be written as a boundary term, we see that it is a conserved quantity.  Furthermore, this form appropriately reduces to the two-dimensional quadrupole conservation law encountered in Reference \onlinecite{elasticity}. In particular, $E^{\,\,iki}_k$ written in terms of the strain tensor is $u^k_k  -  u^{ii}$.  Setting $i=z$ gives  
\begin{equation}
E^{\,\,zkz}_k = u^k_k - u^{zz} = u^{xx} + u^{yy}
\end{equation}  
which is the stretching/compression of the flat surface $S$. Any transverse motion of an edge dislocation line will increase the total volume of the crystal specifically along the cross-section where it moved. This is a conservation law relating dislocation climb to vacancies/interstitials.  Motion of an edge dislocation perpendicular to its direction and Burgers vector will increase this quadrupole flux and therefore stretch the lattice, eventually creating vacancies or interstitials when the stretching becomes large enough.

It must be noted that the formulation of this conservation law makes explicit use of the normal direction to the plane, and therefore breaks rotational invariance. This makes the conservation law in Equation \ref{quadcons} different from those seen earlier, in that the others were written for a general curved 2d surface, while this one is limited to the case of a plane.  Nevertheless, this weaker form of conservation law is sufficient to give rise to the glide constraint of dislocation defects, restricting them to motion within a particular plane.

\subsection{Energetics}

In addition to the mobility restrictions of the line-like lattice defects, an important aspect to consider is their energetics.  We determine the energy associated with defects by working out the interactions between charges of the rank-4 gauge theory.  We will show how these conclusions match up with the expected properties of three-dimensional elasticity theory.  The dominant interaction between charges are electrostatic in nature, and as in conventional electromagnetism, this electrostatic limit can be studied via a potential formulation.  As derived earlier, a static electric tensor configuration can be written as
\begin{equation}
E_{abmn}^{\sigma}=\partial_{m}\partial_{b}\phi_{an}-\partial_{m}\partial_{a}\phi_{nb}+\partial_{a}\partial_{n}\phi_{bm}-\partial_{b}\partial_{n}\phi_{ma}
\end{equation}
for the electrostatic potential $\phi_{ij}$. Plugging $\phi_{ij}$ into the Gauss's law of the theory, and using the relation 
\begin{equation*}
E_{ijkl} = \widetilde{C}^{ijklpqrs}E_{pqrs}^{\sigma} 
\end{equation*}
we obtain a generalized Poisson equation, which we write schematically as:
\begin{equation}
\partial^{4}\phi = \rho
\end{equation}
(We here content ourselves with dimensional analysis, since the precise tensor form of these equations is not as important as their scaling.)  For an isolated fractonic line of charge per unit length $\lambda$ (i.e. a disclination), the potential must then satisfy:
\begin{equation}
\partial^{4}\phi = \lambda \delta^{2}(r)
\end{equation}
Simply by dimensional analysis, we can conclude that the generic solution to this equation is:
\begin{equation}
\phi(r)\sim r^2
\end{equation}
up to logarithmic corrections, where $r$ is the radial coordinate away from the line.  From this form of the potential, we can conclude that:
\begin{equation}
E(r)\sim \partial^2\phi\sim \textrm{constant}
\end{equation}
The total energy per length associated with the disclination line is then:
\begin{equation}
\int d^2r \,E^2 \sim R^2
\end{equation}
where $R$ is some large-distance cutoff, set for example by the system size or by the distance to other disclination lines.  This is indicative of the enormous energy cost necessary to create isolated disclinations within the solid phase.  In practice, therefore, disclination physics is typically only seen on small scales. Nevertheless, they still exist as well-defined topological lattice defects.

We can also make statements about the energetics of dislocation defects, which are simply bound states of two disclinations.  In this case, we can conclude that the electric tensor scales as:
\begin{equation}
E(r)\sim \frac{1}{r}
\end{equation}
and the total energy per length of the dislocation is given by:
\begin{equation}
\int d^2r\,E^2 \sim \log R
\end{equation}
where $R$ is a large-distance cutoff such as the system size.  We therefore see that the energy per length of a dislocation line has a logarithmic behavior exactly mirroring the energy of a point-like dislocation in two-dimensional elasticity theory.

\section{Generalizations}

Up to now, we have focused on a particular example of a gauge theory hosting fractonic line excitations, namely the gauge dual of three-dimensional elasticity theory.  However, it is not difficult to formulate other gauge theories hosting line-like charges with fractonic or subdimensional behavior.  For example, in the model we have studied so far, flux points on a two-dimensional surface behave like the scalar charge theory\cite{sub}, exhibiting conservation of charge and dipole moment.  It would seem plausible that we can construct models where the flux points act like the charges of other symmetric tensor gauge theories.  As an example, we can readily construct a model with flux points having one-dimensional behavior, analogous to the vector charge tensor gauge theory.\cite{sub}

The vector charge theory differs from the scalar charge theory by having only a single derivative in its Gauss's law, leaving an extra index on the charge.  Motivated by this, we consider a modification of the rank-4 gauge theory studied previously, now with only a single derivative in Gauss's law.  We once again consider a rank-4 gauge field $A^{ijk\ell}$ and a rank-4 electric tensor $E^{ijk\ell}$, which are antisymmetric under $i\leftrightarrow j$ and $k\leftrightarrow\ell$, while being symmetric under $(ij)\rightarrow (k\ell)$.  We define the Gauss's law of this theory as:
\begin{equation}
\partial_i E^{ijk\ell} = \rho^{jk\ell}
\end{equation}
with a rank-3 charge tensor $\rho^{jk\ell}$, antisymmetric in the last two indices, which obeys:
\begin{equation}
\partial_j \rho^{jk\ell} = 0
\end{equation}
identically within the entire Hilbert space, forcing the charges to line up into extended objects.  From the Gauss's law, we can also derive the corresponding gauge transformation on the low-energy sector to be:
\begin{equation}
A_{ijk\ell}\rightarrow A_{ijk\ell} + \partial_i\alpha_{jk\ell} - \partial_j \alpha_{ik\ell} + \partial_k\alpha_{\ell ij} - \partial_\ell\alpha_{kij}
\end{equation}
for rank-3 function $\alpha_{jk\ell}$ with arbitrary spatial dependence.  Due to the structure of the gauge transformation, a gauge-invariant magnetic operator must have curls on all indices of $A_{ijk\ell}$.  The lowest order magnetic operator which can be written down in this case is a scalar given by
\begin{equation}
B = \epsilon^{jmn}\epsilon^{abc}\partial_j\partial_c A_{abmn}
\end{equation}
The low-energy Hamiltonian for the theory then generically takes the form:
\begin{equation}
H = \int d^3x\bigg(\frac{1}{2}C^{ijk\ell pqrs}E_{ijk\ell}E_{pqrs} + \frac{1}{2}B^2\bigg)
\end{equation}
leading to a single gapless mode with quadratic dispersion, $\omega\sim k^2$.

We now look for conservation laws on the flux of the line-like charges through two-dimensional surfaces.  As expected, we have conservation of a total flux through an arbitrary surface $S$:
\begin{align}
\begin{split}
\int_S dn_j \rho^{jk\ell} = \int_S dn_j\partial_i E^{ijk\ell} = \int_S dn_j \epsilon^{ijn}\epsilon^{k\ell m}\partial_i V_{nm} \\
=-\oint_{\partial S} ds^n \epsilon^{k\ell m}V_{nm} = \textrm{boundary term}
\end{split}
\end{align}
Note that the choice of contraction between $\rho^{jk\ell}$ and the normal vector is important.  Contraction of $dn$ with either of the last two indices of $\rho$ would not lead to a conserved flux.

Just like the previous rank-4 theory, we can also identify a second conservation law in this theory:
\begin{multline}
\int_S dn_j (\rho^{jk\ell}x^n\epsilon_{k\ell n}) = \int_S dn_j x^b\epsilon^{ija}\partial_i V_{ab} \\
= -\oint_{\partial S} ds^a x^bV_{ab}
\end{multline}
Since this flux moment can be written as a boundary term, it is a locally conserved quantity.  This conservation law is akin to the angular moment conservation law of the vector charge theory.\cite{sub}  It has the consequence that the flux points through the two-dimensional surface will behave as one-dimensional particles.  For example, consider $S$ to be a surface with $z$-directed normal.  It is easy to check that $\rho^{zzx}$ can only move in the $x$ direction, while $\rho^{zzy}$ can only move in the $y$ direction.  In this way, the flux points of this rank-4 tensor gauge theory behave like the gauge charges of the rank-2 vector charge theory.  We expect that, more generally, it will be possible to construct theories with flux points behaving like any of the previously studied tensor gauge theories.  For example, the flux points could have extra higher moment conservation laws, such as conserved quadrupole moment, leading to further restrictions on mobility.

\label{sec:gen}

\section{Conclusions}

In this work, we demonstrated a mapping between elasticity theory of three-dimensional quantum crystals and a rank-4 tensor gauge theory that hosts fractonic line excitations. We showed that the lattice defects of elasticity theory map onto the gauge charges of the rank-4 theory, with the role of wedge disclinations being played by fractonic lines with conserved flux, and that of dislocations by dipoles of flux. Additionally, edge dislocations have the restriction that dislocation climb stretches/compresses the lattice, which creates vacancies/interstitials. We also looked at how the edge and screw dislocations correspond to different bound states of disclinations. Our work opens up the possibility for further dialogue between the fracton and quantum information communities, in that it serves as a reference point for devising new stabilizer codes with loop-like excitations, which have improved error-correction properties.

\section*{Acknowledgments}

We thank Leo Radzihovsky for insightful discussions on elasticity theory and dualities.  MP is supported partially by a Simons investigator award to Leo Radzihovsky, and partially by the NSF Grant 1734006.

\section*{Appendix A: Conservation laws of 2-form gauge theory}

We here review some basics of 2-form gauge theory in three spatial dimensions, supporting conventional line-like excitations, which will be useful for formulating the notion of fractonic line excitations.  Our gauge variable is a rank-2 antisymmetric tensor $A_{\mu\nu}$.  (We take Greek indices to run over space and time, while Latin indices run over space only.)  The theory is invariant under the gauge transformation
\begin{equation}
A_{\mu\nu}\rightarrow A_{\mu\nu}+\partial_\mu \alpha_\nu - \partial_\nu\alpha_\mu
\end{equation}
for arbitrary vector gauge parameter $\alpha_\mu$.  We define a gauge-invariant antisymmetric field tensor as
\begin{equation}
 F^{\mu \nu \rho} = \partial^{\mu}A^{\nu \rho} + \partial^{\nu}A^{\rho \mu} + \partial^{\rho}A^{\mu \nu}.
 \end{equation}
which obeys the equation of motion
\begin{equation}
\partial_{\rho}F^{\mu \nu \rho} = J^{\mu \nu}
\end{equation}
for current tensor $J^{\mu \nu} = - J^{\nu \mu}$.  For our purposes, the most important piece of this equation is the $0i$ component, which gives the Gauss's law in terms of the antisymmetric electric tensor $E_{ij}$ as
\begin{equation}
\partial_i E^{ij} = \rho^j
\end{equation}
where $\rho^j = J^{0j}$.  From this definition, we see that the vector-valued charge $\rho^j$ obeys $\partial_i\rho^i = 0$ exactly, which causes the charges to line up end-to-end in closed string configurations.

We now ask what conservation laws are obeyed by the string-charges $\rho^{i}$ of this theory, analogous to the charge conservation laws of normal gauge theories. For point charges, we can have conservation of quantities such as charge and dipole moment, the latter following from the type of arguments  made in \onlinecite{sub}.  For strings in the rank-2 antisymmetric case, the situation is different:  it is possible to create a small closed string out of the vacuum, so string number is not a conserved quantity. But the number of strings passing through a given two-dimensional surface (i.e. a \textit{flux} of strings) is conserved, and this should be encoded in the one-dimensional edge of that surface.  To see this, we first introduce a rotated electric field, $E^{ij} = \epsilon^{ijk}V_{k}$, in terms of which we have 
\begin{equation}
(\nabla \times V)^{i} = \rho^i
\end{equation}
which is an Ampere-like equation for the quantity $V_{i}$.  By Stokes' theorem, we can now write the flux of $\rho^i$ through an arbitrary surface $S$ as
\begin{equation}
\int_{S}dn^i\rho_i = \oint_{\partial S}ds^iV_i
\end{equation}
We therefore see that the flux of $\rho^i$ through a surface is encoded on the boundary of that surface.  This indicates that a flux line cannot unhook itself from $S$ without crossing the 1d edge, $\partial S$.  No local operator in the interior or $S$ can cause the sudden disappearance of a flux line, meaning that flux is a locally conserved quantity.

Physically, we can understand this bulk-boundary relationship for the charge flux in terms of a duality transformation.  The rank-2 antisymmetric theory is dual to a three-dimensional superfluid, with the strings representing vortices; the number of vortices passing through a 2d surface is determined by the change in the phase angle going around the 1d edge of that surface. Vortex flux conservation in 3d superfluids is given by
\begin{equation*}
\int_{S} dn_{i}\rho^{i} = \oint_{\partial S} ds^{i}\partial_{i}\phi
\end{equation*}
This total vorticity cannot change unless one of the vortices passes through the 1d edge, since a vortex is a line-like topological defect which cannot simply end in the interior of the system.

\section*{Appendix B: Instantons}

In the fracton tensor gauge theory described in the main text, the gauge-invariant magnetic field took the form:
\begin{equation*}
B^{ab} = \epsilon_{jmn} \partial^{j} A^{abmn}
\end{equation*}
If this gauge theory is regarded as noncompact, then there are two separate conservation laws associated with this magnetic field:
\begin{equation}
\int d^{3}x B^{ab} = \textrm{const.}\qquad  \int d^{3}x B^{ab}x_{b} = \textrm{const.}
\end{equation}
For such a noncompact gauge theory, the gapless mode is stable and the charges are deconfined. In a compact gauge theory, on the other hand, $A_{abmn}$ would only be defined modulo some compactification radius, say $2\pi$. In this case, just as in a normal compact $U(1)$ gauge theory, the magnetic flux could slip by units of $2\pi$. Similarly, the moment of flux could slip by units of $2\pi a$, where $a$ is the lattice spacing. Such flux slip events, or instantons, can destabilize the theory, gapping the gauge mode and confining the charges.

Since long-range crystalline order is stable in three dimensions, we expect that elasticity theory should map onto the stable noncompact theory, without instanton processes. But how can we determine this independently, without relying on the known stability of three-dimensional elasticity? This conclusion follows readily from translating the magnetic conservation laws into the original elastic variables, in terms of which we can write:
\begin{equation}
\int d^{3}x\pi_{i} = \textrm{const.}\qquad  \int d^{3}x\epsilon^{ijk}x_{j}\pi_{k} = \textrm{const.}
\end{equation}
which correspond to conservation of linear and angular momentum of the crystal. Importantly, these two conservation laws follow directly from the underlying translational and rotational symmetries of space, which are spontaneously (but not explicitly) broken by the crystal. Thus, for example, any flux-changing instanton event of the gauge theory maps onto a process which does not respect momentum conservation in elasticity theory, i.e. a term which explicitly breaks momentum conservation.  Such terms can arise under specific circumstances, such as via coupling the crystal to a substrate. But for an ordinary crystal in free space, instantons of the gauge theory are ruled out by the underlying translational and rotational symmetries.

\section*{Appendix C: Duality without disclinations}

As discussed in the main text, isolated disclination defects in a crystal have an extremely large energy cost, going as $R^2$ per unit length, where $R$ is some large distance cutoff such as the system size.  As such, these defects can often be disregarded within the low-energy sector.  It therefore seems reasonable to construct an effective low-energy theory purely in terms of the phonons and dislocations, without making reference to disclinations at all.  In this case, we should be able to treat dislocation lines as the fundamental gauge charges, which should have flux points behaving somewhat like the two-dimensional vector charge theory, similar to the model discussed in Section \ref{sec:gen}.  However, that model had a single gauge mode with gapless dispersion, whereas three-dimensional elasticity theory should have three gauge modes with linear dispersion.  As such, we need to add some extra degrees of freedom to obtain the appropriate gauge dual.

As an ansatz, we modify the model of Section \ref{sec:gen} by decreasing the index symmetry of the electric tensor, thereby increasing its number of independent components.  Specifically, we work with a rank-4 electric tensor $E^{ijk\ell}$ which is antisymmetric under $i\leftrightarrow j$ and $k\leftrightarrow\ell$, but does not have any symmetry relating the first and second pair of indices.  With this change in place, we now define our theory using a formally identical Gauss's law to the one used in Section \ref{sec:gen}:
\begin{equation}
\partial_iE^{ijk\ell} = \rho^{jk\ell}
\end{equation}
The corresponding gauge transformation in the low-energy sector is:
\begin{equation}
A_{ijk\ell}\rightarrow A_{ijk\ell} + \partial_i\alpha_{jk\ell} + \partial_j\alpha_{ik\ell}
\end{equation}
The magnetic field operator with fewest derivatives that one can write down takes the form of an antisymmetric tensor:
\begin{equation}
B_{ij} = \epsilon^{k\ell n}\partial_k A_{\ell nij}
\end{equation}
such that the Hamiltonian of the theory takes the generic form:
\begin{equation}
H = \int d^3x\bigg(\frac{1}{2}\tilde{C}^{ijk\ell pqrs}E_{ijk\ell}E_{pqrs} + \frac{1}{2}B^{ij}B_{ij}\bigg)
\end{equation}
In the pure gauge sector, we can replace $E^{ijk\ell}$ by the generic solution to the source-free Gauss's law, $\partial_iE^{ijk\ell} = 0$, which takes the form:
\begin{equation}
E^{ijk\ell} = \epsilon^{ijn}\epsilon^{k\ell m} \partial_n u_m
\end{equation}
The canonical conjugate to $u_i$ takes the form $\pi_i = \epsilon_{ijk}B^{jk}$, such that the Hamiltonian can be written as:
\begin{equation}
H = \int d^3x\bigg(\frac{1}{2}C^{ijk\ell}u_{ij}u_{k\ell} + \frac{1}{2}\pi^i\pi_i\bigg)
\end{equation}
where $u_{ij} = \frac{1}{2}(\partial_iu_j + \partial_ju_i)$ and $C^{ijk\ell} = \epsilon^{imn}\epsilon^{jpq}\epsilon^{krs}\epsilon^{\ell ab}\tilde{C}_{mnpqrsab}$, yielding the standard low-energy Hamiltonian of three-dimensional elasticity theory.

We must also determine the relationship between the tensor charge and the dislocation lines.  To do this, we write a flux of the tensor charges through an arbitrary surface $S$ as:
\begin{align}
\begin{split}
f^{k\ell} &= \int_S dn_j \rho^{jk\ell} = \int_{\partial S} ds^n \epsilon_{nij}E^{ijk\ell} \\
= &\int_{\partial S} ds^n \partial_n \epsilon^{k\ell m}u_m = \epsilon^{k\ell m}\Delta (u_m) = \epsilon^{k\ell m}b_m
\end{split}
\end{align}
We therefore see that the a line charge made from the tensor $\rho^{jk\ell}$ matches up exactly with the dislocation defects of elasticity theory, with charge flux through a two-dimensional surface given by the Burgers vector of the dislocation.  This completes our dual gauge theory formulation of the phonon-dislocation theory.  As we have discussed, this theory does not incorporate disclination defects.  However, for most low-energy purposes, the present theory will provide an accurate treatment.


\begin{thebibliography}{9}
\bibitem{footnote}
Several other gauge theory formulations of elasticity theory are also present in the literature\cite{beekman1,beekman2,gromov}, though the treatment in Reference \onlinecite{elasticity} is the only one to directly map between conventional elasticity theory and a fracton tensor gauge theory.
\end{thebibliography}
\end{document}